\documentclass[twocolumn, prl, superscriptaddress,notitlepage]{revtex4-1}
\usepackage{graphicx}
\usepackage{physics}
\usepackage{color}
\usepackage{amsmath}
\usepackage{amsfonts}
\usepackage{ulem}
\usepackage{xspace}
\usepackage[dvipsnames]{xcolor}
\usepackage{dcolumn}
\usepackage[breaklinks,colorlinks,linkcolor=blue,citecolor=blue,urlcolor=blue]{hyperref}

\newcommand{\Poincare}{Poincar\'e\xspace}
\newcommand{\Mobius}{M\"{o}bius\xspace}
\newcommand{\be}{\begin{equation}}
\newcommand{\ee}{\end{equation}}
\newcommand{\bea}{\begin{eqnarray}}
\newcommand{\eea}{\end{eqnarray}}

\begin{document}
\title{Efimov-like states and quantum funneling effects on synthetic hyperbolic surfaces}

\author{Ren Zhang}
\thanks{They contribute equally to this work.}
\affiliation{School of Physics, Xi'an Jiaotong University, Xi'an, Shaanxi 710049, China}
\affiliation{Department of Physics and Astronomy, Purdue University, West Lafayette, IN, 47907, USA}

\author{Chenwei Lv}
\thanks{They contribute equally to this work.}
\affiliation{Department of Physics and Astronomy, Purdue University, West Lafayette, IN, 47907, USA}

\author{Yangqian Yan}
\affiliation{Department of Physics and Astronomy, Purdue University, West Lafayette, IN, 47907, USA}

\author{Qi Zhou}
\email{zhou753@purdue.edu}
\affiliation{Department of Physics and Astronomy, Purdue University, West Lafayette, IN, 47907, USA}
\affiliation{Purdue Quantum Science and Engineering Institute, Purdue University, West Lafayette, IN 47907, USA}
\date{\today}

\begin{abstract}
Engineering lattice models with tailored inter-site tunnelings and onsite energies could synthesize essentially arbitrary Riemannian surfaces with highly tunable local curvatures. Here, we point out that discrete synthetic \Poincare half-planes and \Poincare disks, which are created by lattices in flat planes, support infinitely degenerate eigenstates for any nonzero eigenenergies. Such Efimov-like states exhibit a discrete scaling symmetry and imply an unprecedented apparatus for studying quantum anomaly using hyperbolic surfaces. 
Furthermore, all eigenstates are exponentially localized in the hyperbolic coordinates, signifying the first example of quantum funneling effects in Hermitian systems. As such, any initial wave packet travels towards the edge of the \Poincare half-plane or its equivalent on the \Poincare disk, delivering an efficient scheme to harvest light and atoms in two dimensions. Our findings unfold the intriguing properties of hyperbolic spaces and suggest that Efimov states may be regarded as a projection from a curved space with an extra dimension. 
\end{abstract}
\maketitle 

{\bf Introduction} 
Quantum simulations have allowed physicists to create a variety of synthetic quantum matters \cite{many-body-review,optics-review,Hartmann_2016,coldatomtopo0,OL-review,coldatomtopo,Iacopo_2020}. However, curved spaces have been less investigated in laboratories, though theorists have predicted novel quantum phenomena in curved spaces that are inaccessible in flat spaces \cite{theorysimulation0,theorysimulation1,theorysimulation2,theorysimulation3,theorysimulation4,theorysimulation5,theorysimulation6,theorysimulation7,theorysimulation8,theorysimulation9,theorysimulation10,theorysimulation11,theorysimulation12}. Despite that it is notoriously difficult to create quantum systems in curved spaces, there have been exciting developments recently. The synthetic cone for cavity photons realized by Simon's group represents a special manifold where the curvatures concentrate at a single point and elsewhere is flat \cite{simon}. The hyperbolic lattice of superconducting circuits used by Houck's group delivers a heptagon tiling of a \Poincare disk with a constant negative curvature, which supplies a new platform to study quantum field theories in curved spaces \cite{hyperdisk,disktheorypra}.  In spite of such progress, a generic scheme is desired to create a curved space with arbitrary distributions of local curvatures.

In this work, we show that engineering unconventional lattice models in flat spaces for both atoms and photons could create the discretized version of a generic Riemannian surface. As examples, we show how to create a discrete synthetic \Poincare half-plane and a discrete synthetic \Poincare disk using two-dimensional lattices in flat planes. A profound property of such hyperbolic surfaces is that they support an infinite number of Efimov-like states, which exhibit a discrete scaling symmetry. It is known that the discrete scaling symmetry, which originates from quantum anomaly breaking the continuous symmetry, is a characteristic property of the Efimov states \cite{Efimov}, a celebrated three-body bound state of fundamental importance in atomic and nuclear physics \cite{Braaten-review,Naidon_review,Chris-review}. Strikingly, similar effects 
have also been discovered in graphene and other topological materials \cite{graphene,ultraquantumlog}, suggesting that Efimov physics may not be unique to scattering problems in few-body systems but exist in a broad class of systems.

Here, the Efimov-like eigenstates unfold the importance of the underlying symmetry of the hyperbolic surfaces. Imposing a boundary condition naturally breaks the continuous scaling symmetry to a discrete one, and quantum anomaly rises \cite{anomaly}. Moreover, a rigorous mapping can be established between the Schr\"odinger equation on a Poincare half-plane (or a Poincare disk) and the hyper-radial equation of Efimov states, despite that these two equations concern systems in distinct dimensions. If we utilize hyperbolic coordinates, a quantum funneling effect becomes evident. All eigenstates are exponentially localized near the funneling mouth, the edge of the \Poincare half-plane or its equivalent on the \Poincare disk. As a result, in a quench dynamics, such as suddenly changing a flat space to a hyperbolic surface, any initial wave packets must travel towards the funneling mouth. A similar effect has recently been found in one-dimensional systems, where non-Hermiticity enforces the localization of all eigenstates at one end \cite{topofunneling,skineffect}. Here, our results of two-dimensional hyperbolic surfaces signify the first example of quantum funneling effects  in Hermitian systems and demonstrate the power of synthetic curved spaces as a new tool to manipulate quantum dynamics such as harvesting light and atoms.

Different from the scheme used in Houck's group \cite{hyperdisk}, which implements a spatially non-uniform distribution of lattice sites with a uniform inter-site tunneling, we consider spatial-uniformly distributed lattice sites. Since uniformly distributed lattice sites naturally exist in many systems, our approach does not require to fine-tune the locations and the inter-site tunnelings simultaneously. More importantly, it provides experimentalists with a simple recipe to dynamically control the metric of the synthetic curved spaces without changing locations of lattice sites. For instance, tuning the inter-site tunneling and onsite energies allows experimentalists to study a new type of quench dynamics when the metric of the spaces suddenly changes. We note that a similar scheme was considered as a proposal for studying analog gravitational waves in optical lattices \cite{szpakcurved}.

{\bf Results}

{\it Lattice models for Riemann surfaces.}
We consider a two-dimensional Riemann surface, the metric of which can be conformally mapped to a Euclidean metric, i.e., 
\begin{align}
ds^2=f(x,y)({\rm d}x^2+{\rm d}y^2). \label{metric}
\end{align}
 The energy functional of a non-relativistic particle on this surface is written as \cite{gutzwiller_chaos}
\begin{align}
\begin{split}
H=& \int dxdy \sqrt{g}\left(-\frac{1}{\sqrt{g}}\sum_{i=x,y}\Psi^*\partial_i\sqrt{g}g^{ii}\partial_{i}\Psi-\frac{\kappa}{4}|\Psi|^2\right),
\end{split}
\end{align}
where $g_{xx}=g_{yy}=f(x,y)$, $g=f(x,y)^2$ and $g^{xx}=g^{yy}=1/f(x,y)$, and $-\kappa$ is the Gaussian curvature. We adopt the unit $\hbar=2m=1$ hereafter, where $m$ is the single particle mass.
The wavefunction $\Psi(x,y)$ satisfies the Schr\"odinger equation, 
\begin{align}
\label{general-curved-model}
\left(-\frac{1}{\sqrt{g}}\sum_{i=x,y}\partial_i\sqrt{g}g^{ii}\partial_{i}-\frac{\kappa}{4}\right)\Psi(x,y)=E\Psi(x,y),
\end{align}
and the normalization condition reads
\begin{align}
\int dxdy \sqrt{g} |\Psi(x,y)|^2=1.
\end{align}

\begin{figure}
\centering
\includegraphics[width=0.45\textwidth]{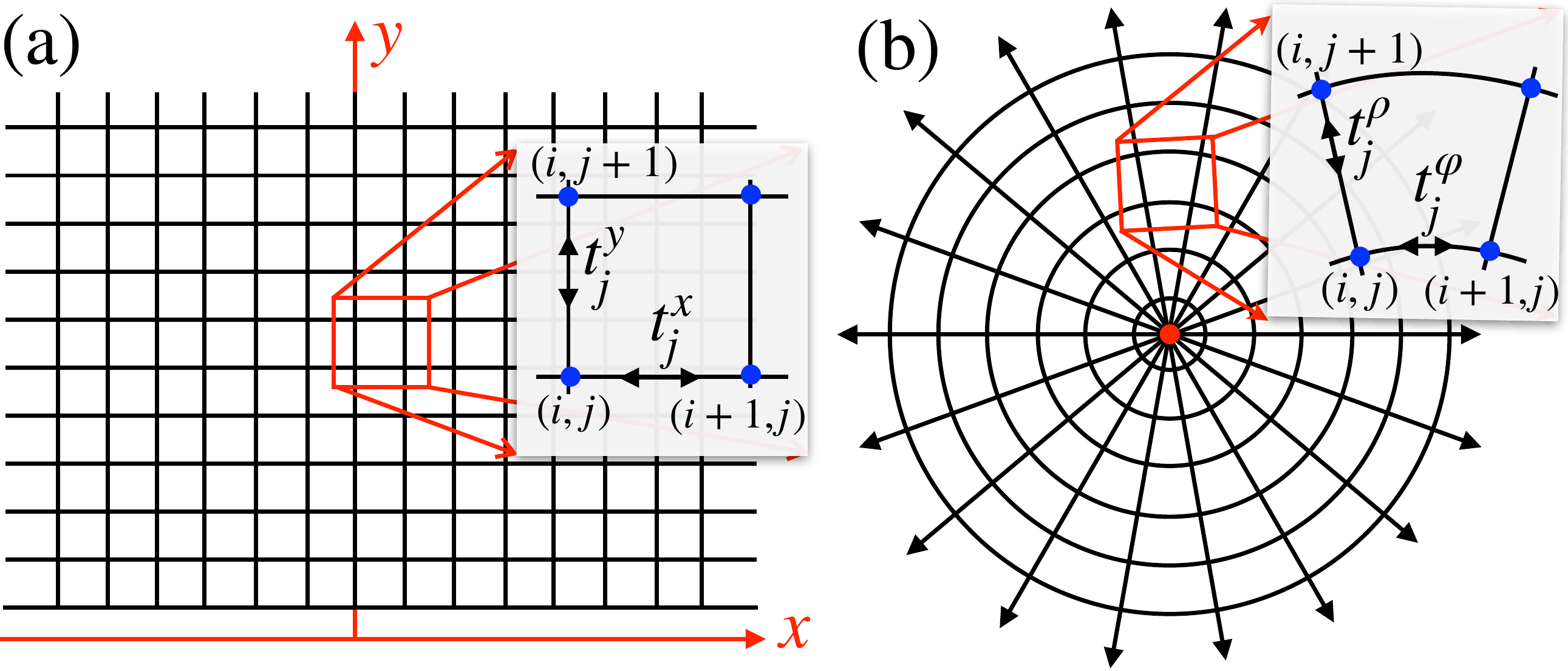} 
\caption{ Discretization of the \Poincare half-plane (a) and \Poincare disk (b) using spatial-uniformly distributed lattice sites. The inter-site tunnelings and on-site energies in both cases are non-uniform (see text). \label{cartoon}}
\end{figure}

For a \Poincare half-plane, $f(x,y)=1/(\kappa y^2)$ with $\kappa>0$ \cite{chaosonpseudosphere}, and the Gaussian curvature, therefore, is negative. Discretizing the continuous model on uniformly distributed lattice sites, the lattice model is given by 
\begin{align}
 \label{lattice-half-plane}
 \hat{H}_{\rm p}=\sum_{i=-\infty}^{\infty}\sum_{j=0}^{\infty}a^{\dagger}_{i,j}&\Big[t^x_{j}a_{i+1,j}+t^y_{j}a_{i,j+1}+u_{i,j}a_{i,j}\Big]+{\rm h.c.},
\end{align}
where $t^x_{j}=-\kappa j^{2}$, $t^y_{j}=-\kappa j(j+1)$ and $u_{i,j}=(4j^{2}-1/4)\kappa$. $a_{i,j}$ and $a^{\dagger}_{i,j}$ denote the annihilation and creation operator on site $(i,j)$, respectively. 
$i(j)$ is the lattice site index along the $x(y)$-direction. Since we consider the upper half-plane, $j$ starts from 0.
 The non-uniform tunneling along the $x$-direction is proportional to $j^{2}$, and that along the $y$-direction is proportional to $j(j+1)$ as shown in Fig. \ref{cartoon}(a). 
 This is a direct consequence of an intrinsic property of the \Poincare half-plane. If we fixed the Euclidean distance between two points, their distance on the \Poincare half-plane decreases with increasing $y$. The solution of the lattice model in Eq.(\ref{lattice-half-plane}), $\varphi_{i,j}$, directly provides us with the wavefunction on the \Poincare half-plane via $\Psi(x_i, y_j)=\varphi_{i,j} g_{i,j}^{-1/4}$  (Supplementary Note 1).

A \Poincare half-plane can be mapped to another hyperbolic surface, the \Poincare disk, whose metric is written as \cite{chaosonpseudosphere}
\begin{align}
 ds^2=\frac{4}{(1-\kappa\rho^2)^2}{\rm d}\rho^2+\frac{4\rho^2}{(1-\kappa\rho^2)^2}{\rm d}\theta^2,\label{metricd}
\end{align}
where $(\rho, \theta)$ are the polar coordinates. Whereas it can be rewritten in the form of Eq.(\ref{metric}), here, we make explicit use of Eq.(\ref{metricd}). To this end, we consider a discrete lattice as shown in Fig. \ref{cartoon}(b). Discretizing the Schr\"odinger equation in polar coordinates leads to a lattice model
\begin{align}
\label{lattice-disk}
\hat{H}_{\rm d}=&\sum_{i,j}a^{\dagger}_{i,j}\left[t^\rho_{j} a_{i,j+1}+t^\varphi_{j} a_{i+1,j}+u_{i,j}a_{i,j}\right]+{\rm h.c.},
\end{align}
where $t^\rho_{j}=- j\sqrt{\frac{t(j+1)t(j)}{j(j+1)}}$, $t^\varphi_{j}=- t(j)/(j^{2}\alpha^{2})$ and $u_{i,j}=\frac{\kappa}{4}-\frac{2j-1}{j}t(j)-\frac{2t(j)}{j^{2}\alpha^{2}}$ with $t(j)=(1-j^{2}a^{2}\kappa)^{2}/(4a^{2})$. $i$ and $j$ are the lattice site  
indices along the azimuthal and radial direction, respectively. $a$ and $\alpha$ are the corresponding lattice constants along these two directions. 
Similar to the half-plane model in Eq.(\ref{lattice-half-plane}), the inter-site tunneling and the onsite energies here are also non-uniform.

{\it Realizations in experiments.}
There are a variety of systems in quantum optics and ultracold atoms to realize the aforementioned lattice models for both the \Poincare half-plane and the \Poincare disk. Since the connectivity between superconducting circuits or optical resonators, as well as the onsite energy in each circuit  or resonator, are highly tunable \cite{circuitQED1,circuitQED2,opticalresonator1,opticalresonator2,opticalresonator3,opticalresonator4}, it is a natural choice to explore these hyperbolic surfaces, as well as other curved spaces, in quantum optics. In parallel, a digit micromirror device can be used to design the landscape of the external potentials for atoms \cite{DMD1,DMD2,DMD3,DMD4}. Thus, in ultracold atoms, a lattice model with desired inter-site tunnelings and onsite energies, such as those shown in Eq.(\ref{lattice-half-plane}) and Eq.(\ref{lattice-disk}), could be realized. It is also possible to realize a hyperbolic half-plane using ordinary optical lattices. Similar to the scheme realizing the Harper-Hofstadter model, a field gradient tilts the lattice potential and thus suppresses the bare tunnelings of atoms \cite{Harper,Hofstadter}. Adding external lasers to induce photon-assisted tunnelings, non-uniform tunnelings can be achieved by controlling either the spatially variant amplitude of the laser or the site-dependent two-photon detuning. Alternatively, synthetic dimensions could be implemented. For instance, in the momentum space lattice, the coupling strength between different momentum states can be independently controlled \cite{LENS-ribbons,Stuhl1514,Seoul,Purdue,UIUC}. Thus, the momentum space lattice, or similar ones from other synthetic dimensions, could also be an appropriate platform to study the synthetic curved spaces. 

\begin{table*}[t]
\centering
\begin{tabular}{|>{\centering\arraybackslash}m{3cm}|>{\centering\arraybackslash}m{3cm}|>{\centering\arraybackslash}m{3cm}|>{\centering\arraybackslash}m{3cm}|>{\centering\arraybackslash}m{4cm}|} 
\hline 
\Poincare half-plane & Position $y$ & Kinetic energy in the $x$-direction $k_x^2$
& Energy/Gaussian curvature $\sqrt{E/\kappa}$ & Scaling between $k_x^2$ $k^2_{x,n-1}=e^{2\pi\sqrt{\kappa/E}}k^2_{x,n}$ \\
\hline 
Efimov physics & Hyper radius $R$ &Binding energy $\tilde{E}_b$ 
& Scaling factor $s_0$ & Scaling between $\tilde{E}_{b}$ $\tilde{E}_{b,n-1}=e^{2\pi/s_{0}}\tilde{E}_{b,n}$ \\
\hline
\end{tabular}
\caption{Comparisons between the Efimov-like states in the \Poincare half-plane and the Efimov states in three-body systems. }
\label{table}
\end{table*}

\begin{figure}[t]
\centering
\includegraphics[width=0.45\textwidth]{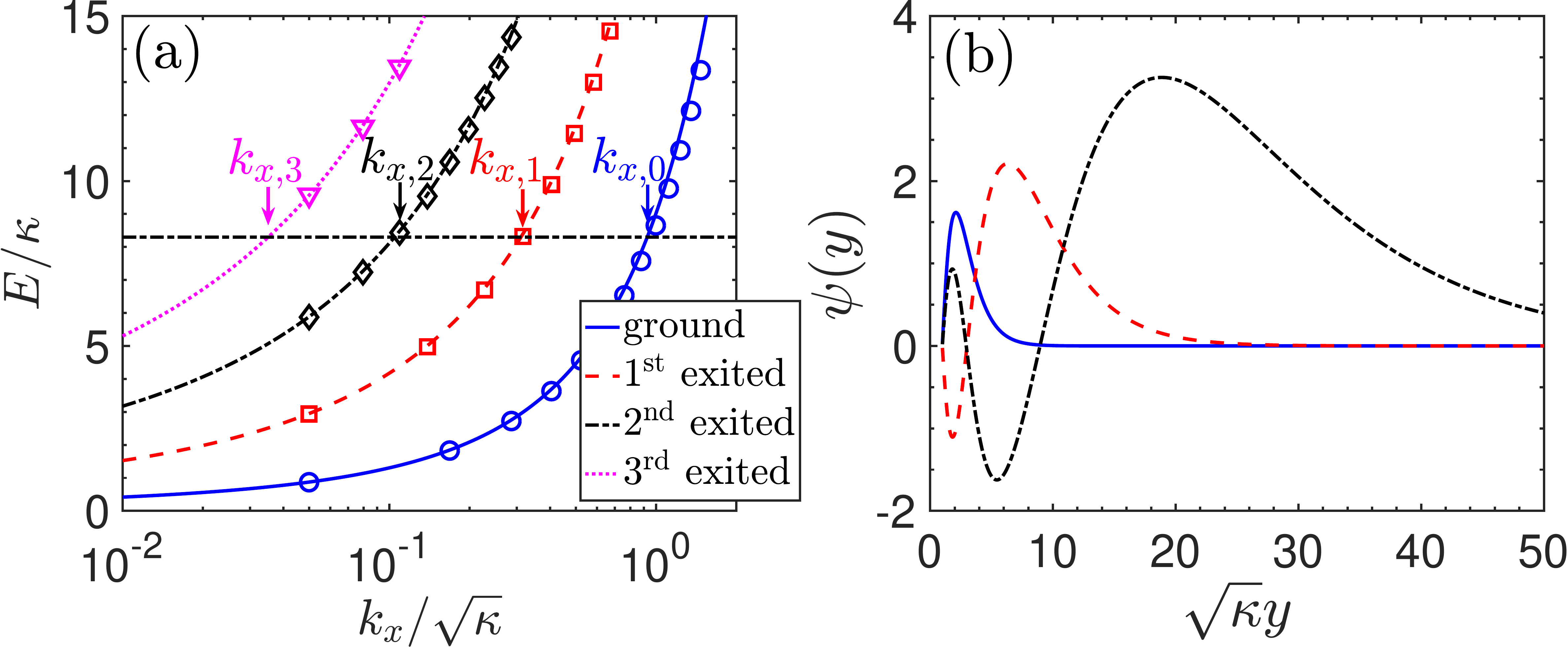} 
\caption{(a): Comparisons of eigenenergies of the continuous model (curves) and the discretized lattice model (markers). $k_{x,n-1}/k_{x,n}=e^{\pi\sqrt{\kappa/E}}$. (b): Wavefunctions of three  eigenstates in the infinitely degenerate manifold for $E/\kappa=8.5$. The short-range boundary condition is $\psi(1/\sqrt{\kappa})=0$.
\label{comparision}}
\end{figure}

{\it Efimov-like states.}
We now turn to the eigenstates in a \Poincare half-plane. Using the metric tensor of the \Poincare half-plane, one finds the stationary Schr\"odinger equation 
\begin{align}
\label{half-plane-model}
-\kappa\left[y^{2}\left(\frac{\partial^{2}}{\partial x^{2}}+\frac{\partial^{2}}{\partial y^{2}}\right)+\frac{1}{4}\right]\Psi(x,y)=E\Psi(x,y),
\end{align}
where $y>0$, and $\int_{0}^{\infty}\frac{dy}{\kappa y^2}\int_{-L}^{L}{dx}\Psi^{*}(x,y)\Psi(x,y)=1$.
Motions along the $x$ and $y$-directions in the \Poincare half-plane are separable, 
and the wavefunction can be factorized as $\Psi(x,y)=\phi(x)\psi(y)$, where $\phi(x)=\frac{1}{\sqrt{L}}e^{ik_xx}$ and $\psi(y)$ satisfies
\begin{align}
\label{efimov-e}
\frac{\partial^{2}}{\partial y^{2}}\psi(y)+\frac{1/4+E/\kappa}{ y^{2}}\psi(y)=\epsilon\psi(y).
\end{align}
$\epsilon=k_x^2$ is the kinetic energy in the $x$-direction. Eq.(\ref{efimov-e}) is scaling invariant and is identical to the hyper-radial equation of three identical bosons with resonant pairwise interactions \cite{Efimov}, if we use the mapping, $y\rightarrow R$, $E/\kappa\rightarrow s_0^2$, and $\epsilon\rightarrow -\tilde{E}_b$. $R$ is the hyper radius, $s_0\approx1.00624$ is the discrete scaling factor. $\tilde{E}_b$ is the binding energy of an Efimov state. In Efimov physics, for a given $s_0$, there are infinite numbers of bound states, whose energies satisfy $\tilde{E}_{b,n-1}=e^{2\pi/s_0}\tilde{E}_{b, n}$. As such, the energy spectrum in the \Poincare half-plane has a unique feature. When $E=0$, there is only one eigenstate with $k_x=0$. For finite $E>0$, there are infinitely degenerate {\it Efimov-like} eigenstates, since an infinite number of $\epsilon$, or equivalently, the momentum in the $x$ direction, is allowed as the solution to Eq.(\ref{efimov-e}). These degenerate states obey a discrete scaling law in the same manner as Efimov states once a boundary condition is applied.

To be specific, the wavefunction satisfying Eq.(\ref{efimov-e}) reads $\psi(y)\propto\sqrt{y}K_{i\sqrt{E/\kappa}}(\sqrt{\epsilon}y)$ where $K_{a}(y)$ denotes the modified Bessel function of the second kind. In the limit of $\epsilon\to0$, the wavefunction could be rewritten in a more insightful form $\psi(y)\propto\sqrt{y}\cos(\sqrt{E/\kappa}\ln\frac{\sqrt{\epsilon}y}{2}+\theta)$ with $\theta=\arg[\Gamma(-i\sqrt{E/\kappa})]$. 
By imposing the boundary condition, $\psi(y_{0})=0$, $k_x$ and $\epsilon$ become quantized for a given $E$, and the continuous scaling symmetry reduces to a discrete one, $\epsilon_{n-1}=e^{2\pi\sqrt{\kappa/E}}\epsilon_{n}$, or equivalently, $k_{x,n-1}=e^{\pi\sqrt{\kappa/E}}k_{x,n}$. In other words, quantum anomaly occurs. Correspondingly, $\psi_n(y)\propto \psi_{n-1}(ye^{\pi\sqrt{\kappa/E}})$, i.e., the dilation $y\to ye^{\pi\sqrt{\kappa/E}}$ transforms the $(n-1)$-th eigenstate in the infinitely degenerate manifold to the $n$-th one, as shown in Fig.~\ref{comparision}. Here, the scaling factor is determined by the eigenenergy $E$ and the Gaussian curvature $\kappa$, and thus can be continuously tuned, unlike Efimov states, whose scaling factor, $s_0$, is fixed by the mass ratios and particle statistics. When $\epsilon=0$, the wavefunction becomes $\psi_{\infty}(y)\propto\sqrt{y}\sin(\sqrt{E/\kappa}\ln\frac{y}{y_{0}})$, which is invariant under the dilation $y\to ye^{\pi\sqrt{\kappa/E}}$. 

\begin{figure}[t]
\centering
\includegraphics[width=0.45\textwidth]{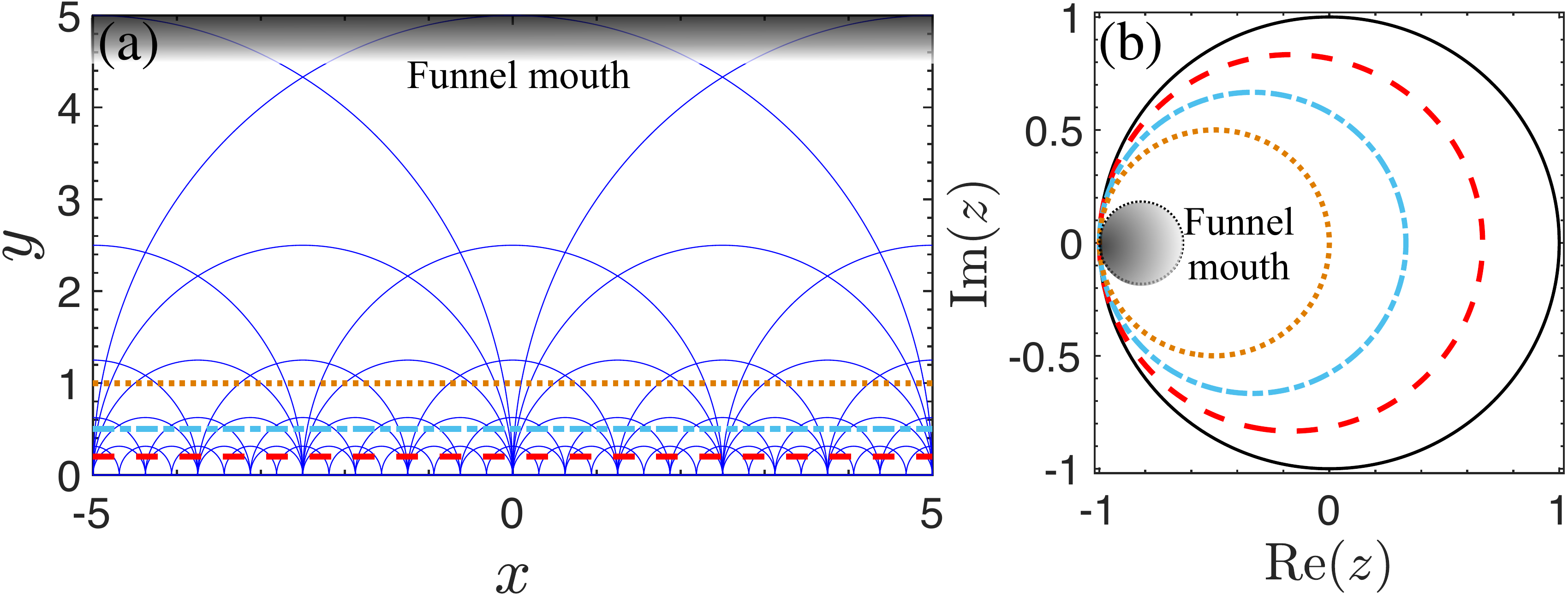} 
\caption{Boundaries $y=y_{0}$ on the \Poincare half-plane (a) are mapped to a series of horocycles on the \Poincare disk (b) by a \Mobius transformation in Eq. (\ref{mobius}). 
$y_{0}= 0.2$ (red dashed), $0.5$ (cyan dash-dotted) and $1$(yellow dotted). The blue curves denote the geodesic of \Poincare half-plane. 
Shaded regions denote funneling mouths on these two hyperbolic surfaces. \label{halfplane-disk}}
\end{figure}
Since the \Poincare half-plane and the \Poincare disk are related by a Cayley transformation \cite{chaosonpseudosphere}
\begin{equation}
\label{mobius}
\rho e^{i\theta}=-\frac{x+i(y-1)}{x+i(y+1)},
\end{equation}
the results of a \Poincare half-plane allow us to directly obtain the eigenstates on the \Poincare disk. We need to emphasize that for such mapping to work, the boundary conditions should also be transformed correspondingly. As shown in Fig. \ref{halfplane-disk}, the boundaries $y=y_{0}$
on the \Poincare half-plane are transformed to horocycles, which are curves with perpendicular geodesics converging asymptotically in the same direction, on the \Poincare disk. Only under this particular boundary condition, the Efimov-like states emerge on the \Poincare disk. For other boundary conditions, results will be modified (Supplementary Note 2).

As for the discrete lattice models, we numerically solve them and compare the numerical results with the analytical ones of the continuous model. Using the \Poincare half-plane as an example, we show in Fig. \ref{comparision} that a good agreement, as expected, confirms the validity of the lattice models as discrete versions of the hyperbolic spaces.

If we take a closer look at the eigenstates on these two hyperbolic surfaces, we note that all eigenstates are localized in the hyperbolic coordinates. Since the distance from $y_0$ to $y$ on the \Poincare half-plane is
\begin{align}
s=\int_{y_0}^y dy\frac{1}{\sqrt{\kappa}y}=\frac{1}{\sqrt{\kappa}}\ln\frac{y}{y_0}, 
\end{align}
the eigenstates can be rewritten as $\psi_n(s)\propto\sqrt{y_{0}}e^{\sqrt{\kappa}s/2}K_{i\sqrt{E/\kappa}}(\sqrt{\epsilon_n}y_{0}e^{\sqrt{\kappa}s})$. When $\epsilon=0$, which corresponds to the disassociation threshold of the Efimov states, the wavefunction can be simplified as $\psi_\infty(s)\propto\sqrt{y_{0}}e^{\sqrt{\kappa}s/2}\sin(\sqrt{E}s)$. Thus, an observer living in the Euclidean space could find that all eigenstates are exponentially localized around $s=\infty$. Such a localization at infinity is precisely due to the particular behavior of the metric of the hyperbolic space, which is proportional to $1/y^2$. A finite $\epsilon$ introduces an extra effective  potential in the $y$-direction to compensate the localization. As such, for any finite $\sqrt{\epsilon}$, any eigenstates decay to zero when $s\to\infty$. Nevertheless, in a length scale that is much smaller than the characteristic length scale $\sim1/\sqrt{\epsilon}$ of the effective potential, the tendency of the localization towards infinity remains.

{\it Quantum funneling effect.}
Since all eigenstates are exponentially localized, we refer to such a phenomenon as a quantum funneling effect, reminiscent of the funneling effects in certain one-dimensional non-Hermitian systems, where all eigenstates are also localized \cite{topofunneling,skineffect}. In these non-Hermitian systems, the chiral tunnelings force eigenstates to concentrate at one end of a one-dimensional chain. Here, we are considering Hermitian systems where chiral tunnelings are absent. It is the metric of a curved space that induces the quantum funneling in two dimensions. On the Poincare half-plane, the funneling mouth locates at $y=\infty$. Again, due to the mapping between the \Poincare disk and the \Poincare half-plane, the same phenomenon of the localization of all eigenstates also occurs on the \Poincare disk. The only quantitative difference is that the funneling mouth is located around a particular point on the \Poincare disk, as shown by the shaded regions in Fig. ~\ref{halfplane-disk}. The locations of funneling mouths on the \Poincare plane and \Poincare disk can be continuously changed by implementing appropriate boundary conditions (Supplementary Note 3).

\begin{figure}[t]
\centering
\includegraphics[width=0.4\textwidth]{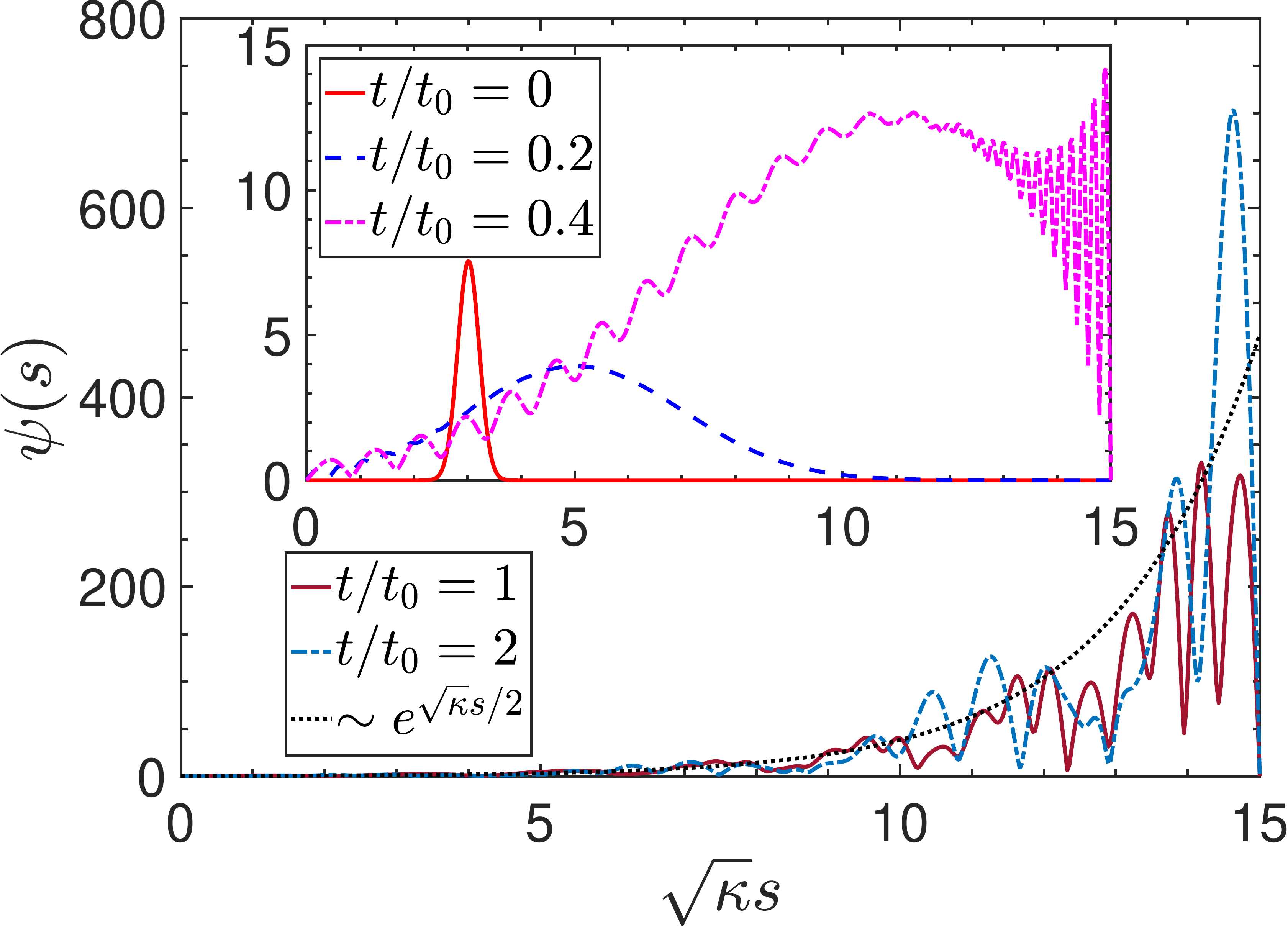} 
\caption{Quantum funneling effects in the \Poincare half-plane. A localized wave packet prepared at $t=0$ (the red solid curve in the inset) smears and travels to the funneling mouth at the higher end of $s$ as time goes by. In the long-time limit, the wave packet localizes at the boundary with an exponential envelope (black dotted curve). We have $t_{0}=2m/(\hbar^{2}\kappa)$ as the time unit. Labels of the inset are the same as the main figure, showing short-time dynamics.\label{dynamics}}
\end{figure}

The quantum funneling effect has a profound impact on quantum dynamics. Any initial states must travel towards the funneling mouth. 
In Fig.~\ref{dynamics}, we demonstrate the quantum funneling effect using a wave packet in the \Poincare half-plane, which has a vanishing $k_x$ and is a Gaussian in the $y$-direction at $t=0$. This initial state could be regarded as the ground state in a flat plane with harmonic confinement in the $y$-direction. At $t=0$, we suddenly remove the harmonic confinement and change the metric to that of a \Poincare half-plane. This represents a new type of quench dynamics of suddenly curving a flat plane to a hyperbolic surface. Whereas it is difficult to perform such a task using conventional apparatuses, in our scheme of discretized lattices, experimentalists just need to quench inter-site tunnelings and onsite energies without physically moving each lattice site.

Since $k_x$ is a good quantum number, we focus on the dynamics in the $y$ direction. In the hyperbolic coordinates, 
the normalization of wavefunction $\psi(s)$ reads $\frac{1}{\kappa}\int \frac{ds}{e^{s}}\psi^{*}(s)\psi(s)= 1$. As shown in Fig.~\ref{dynamics}, while the wave packet expands, it travels to the higher end of $s$ where the funneling mouth locates. In the long-time limit, the wave packet localizes at the boundary with an exponential envelope, which is analog of the funneling on non-Hermitian \cite{topofunneling}. In the numerics for $k_x=0$, we have included a cut-off at large $y$. The wave packet would otherwise continuously move towards infinity. For any finite $k_x\neq 0$, such a large distance cut-off naturally exists because of the extra effective confinement in the $y$-direction. For both the vanishing and finite $k_x$, to unfold the funneling effects using the discrete lattices, one just needs to implement the aforementioned mapping between the wavefunction in the lattice model and that in the continuous space and then apply the coordinate transformation $y\rightarrow s$. Since our results apply to both optical lattices and superconducting circuits, quantum funneling effects discussed here provide experimentalists with an efficient tool to harvest atoms and light in two dimensions.

{\bf Discussions}

In addition to the previously discussed intriguing single-particle physics, interaction effects can also be taken into account on the synthetic hyperbolic surfaces. For example, the GP equation for interacting bosons on a curved surface is written as,
\begin{align}
i\partial_{t}\Psi(x,y;t)=&\left(-1/\sqrt{g}\sum_{i=x,y}\partial_i\sqrt{g}g^{ii}\partial_{i}-\kappa/4\right)\Psi(x,y;t)\nonumber\\
&+u N|\Psi(x,y;t)|^{2}\Psi(x,y;t),
\end{align}
where $u$ is the interaction strength and $N$ is the particle number.
To incorporate the interaction effect in the lattice model, we discretize the interacting Hamiltonian on uniformly distributed lattice sites, leading to 
a site-dependent interaction $H_{\rm int}=\sum_{i,j}U_{i,j} a^\dagger_{i,j}a^\dagger_{i,j}a_{i,j}a_{i,j}$.
Here $U_{i,j}=u/(ab\sqrt{g_{i,j}})$ with $a,b$ denoting the lattice constants.
For the \Poincare half-plane and the \Poincare disk, $U_{i,j}=u\kappa j^2$ and $U_{i,j}=u(1-j^{2}a^{2}\kappa)/(4ja^{2}\alpha)$, respectively.
In superconducting circuits, such non-uniform onsite interactions can be realized by individually controlling the anharmonicity of each circuit \cite{circuitQED1,circuitQED2}. In ultracold atoms, either optical or magnetic Feshbach resonances could control the local scattering length to realize a desired onsite interaction energy \cite{OpFR1,OpFR2,OpFR3,SpatialFR}.

Whereas this work focuses on the intriguing properties of hyperbolic surfaces, our results also shed new light on Efimov physics. Despite the distinct microscopic physics in a three-body problem and the hyperbolic surfaces, Table \ref{table} is very suggestive. The gauge theory/gravity dual has told us that the energy could be regarded as an extra dimension in the renormalization group (RG) language \cite{adscft_review}. Table~\ref{table} thus raises an interesting question of whether 
Efimov states can be viewed as projections from a hyperbolic space with an extra dimension, the energy of the Efimov state, $\tilde{E}_b$, which corresponds to $k_x^2$ in the \Poincare half-plane. The dissociation of the Efimov states when $\tilde{E}_b$ approaches the threshold then may be regarded as a quantum funneling effect on a hyperbolic surface. 

In addition to hyperbolic surfaces, our scheme can be generalized to higher dimensions. Engineering non-uniform tunnelings and onsite energies in synthetic dimensions will allow experimentalists to access curved spaces in $D>3$. We hope that our work will inspire interest to study not only synthetic curved spaces but also connections between Efimov physics and curved spaces. 

{\bf Acknowledgements}

This work is supported by the Air Force Office of Scientific Research under award number FA9550-20-1-0221, DOE DE-SC0019202, W. M. Keck Foundation, and a seed grant from Purdue Quantum Science and Engineering Institute. 
R.Z. is supported by NSFC (Grant No.11804268) and the National Key R$\&$D Program of China (Grant No. 2018YFA0307601).

{\bf Data availability} The data that support the findings of this study are available from the corresponding
authors upon request.

{\bf Author contributions}
R.Z. and C.L. conducted analytical and numerical calculations with inputs from Y. Y. and Q.Z.  Q.Z. conceived and supervised the project. All the authors contributed to the writing of the manuscript.

{\bf Competing interests}
The authors declare no competing interests.

\bibliographystyle{apstest}
\bibliography{Efimov_NC_v4.bib}

\clearpage

\pagebreak
\widetext
\begin{center}
\textbf{\large{}Supplementary Information: Efimov-like states and quantum funneling effects on synthetic hyperbolic surfaces'}
\end{center}
\setcounter{equation}{0}
\setcounter{figure}{0}
\setcounter{table}{0}
\makeatletter
\renewcommand{\theequation}{S\arabic{equation}}
\renewcommand{\thefigure}{S\arabic{figure}}


\subsection{Supplementary Note 1: A uniform discretization of continuous models on Riemann surfaces}
The energy functional on a Riemann surface is written as
\begin{align}
H=&-\frac{\hbar^{2}}{2m}\int \sqrt{g}dxdy\Psi^{*}(x,y)\frac{1}{\sqrt{g}}\left[\frac{\partial}{\partial x}\sqrt{g}g^{xx}\frac{\partial}{\partial x}+\frac{\partial}{\partial y}\sqrt{g}g^{yy}\frac{\partial}{\partial y}\right]\Psi(x,y)+\frac{\hbar^{2}\kappa}{8m}\int \sqrt{g}dxdy|\Psi(x,y)|^{2},
\end{align}
where $g^{xx}=g_{xx}^{-1}$ and $g^{yy}=g_{yy}^{-1}$ are metric tensor elements and $g=g_{xx}g_{yy}$. The normalization of the wave function, $\Psi(x,y)$, is written as
\begin{align}
\int \sqrt{g}dxdy|\Psi(x,y)|^{2}=1.
\end{align}
Using a uniform discretization, $x_{i+1}-x_{i}=a,\ y_{j+1}-y_{j}=b$, the energy functional then becomes
\begin{align}
\begin{split}
H=&-\frac{\hbar^{2}}{2ma^{2}}\sum_{i,j}\sqrt{g_{i,j}}\left[g_{i,j}^{xx}\left(\Psi^{*}_{i+1,j}\Psi_{i,j}+\Psi^{*}_{i,j}\Psi_{i+1,j}\right)+\frac{a^{2}}{b^{2}}g_{i,j}^{yy}\left(\Psi^{*}_{i,j+1}\Psi_{i,j}+\Psi^{*}_{i,j}\Psi_{i,j+1}\right)\right]\\
&+\sum_{i,j}\sqrt{g_{i,j}}\left[\frac{\hbar^{2}}{2ma^{2}}\left(\frac{\sqrt{g_{i-1,j}}}{\sqrt{g_{i,j}}}g_{i-1,j}^{xx}+g_{i,j}^{xx}+\frac{a^{2}}{b^{2}}\frac{\sqrt{g_{i,j-1}}}{\sqrt{g_{i,j}}}g_{i,j-1}^{yy}+\frac{a^{2}}{b^{2}}g_{i,j}^{yy}\right)-\frac{\hbar^{2}\kappa}{8m}\right]\Psi_{i,j}^{*}\Psi_{i,j}
\end{split}
\end{align}
where $\Psi_{i,j}=\sqrt{ab}\Psi(ia,jb)$. Here $a$ and $b$ denote the lattice constant along $x$ and $y$-direction, respectively.
The normalization is written as
\begin{align}
\label{norm1}
\sum_{i,j}\sqrt{g_{i,j}}\Psi_{i,j}^{*}\Psi_{i,j}=1.
\end{align}
The coefficient $\sqrt{g_{i,j}}$ in the normalization Eq.(\ref{norm1}) can be absorbed by defining
\begin{align}
\varphi_{i,j}=g^{1/4}_{i,j}\Psi_{i,j}.
\end{align}
As a result, the energy functional is rewritten as
\begin{align}
\begin{split}
H=&-\frac{\hbar^{2}}{2ma^{2}}\sum_{i,j}\left[\frac{\sqrt{g_{i,j}}g_{i,j}^{xx}}{(g_{i+1,j}g_{i,j})^{1/4}}\left(\varphi^{*}_{i+1,j}\varphi_{i,j}+\varphi^{*}_{i,j}\varphi_{i+1,j}\right)+\frac{a^{2}}{b^{2}}\frac{\sqrt{g_{i,j}}g_{i,j}^{yy}}{(g_{i,j+1}g_{i,j})^{1/4}}\left(\varphi^{*}_{i,j+1}\varphi_{i,j}+\varphi^{*}_{i,j}\varphi_{i,j+1}\right)\right]\\
&+\sum_{i,j}\left[\frac{\hbar^{2}}{2ma^{2}}\left(\frac{\sqrt{g_{i-1,j}}}{\sqrt{g_{i,j}}}g_{i-1,j}^{xx}+g_{i,j}^{xx}+\frac{a^{2}}{b^{2}}\frac{\sqrt{g_{i,j-1}}}{\sqrt{g_{i,j}}}g_{i,j-1}^{yy}+\frac{a^{2}}{b^{2}}g_{i,j}^{yy}\right)-\frac{\hbar^{2}\kappa}{8m}\right]\varphi_{i,j}^{*}\varphi_{i,j}
\end{split}
\end{align}
with the normalization $\sum_{i,j}\varphi_{i,j}^{*}\varphi_{i,j}=1$. The second quantized Hamiltonian is written as
\begin{align}
\label{2ndH}
\hat{H}=&\sum_{i,j}\left[t_{i,j}^{x}\left(a^{\dagger}_{i+1,j}a_{i,j}+a^{\dagger}_{i,j}a_{i+1,j}\right)+t_{i,j}^{y}\left(a^{\dagger}_{i,j+1}a_{i,j}+a^{\dagger}_{i,j}a_{i,j+1}\right)+u_{i,j}a_{i,j}^{\dagger}a_{i,j}\right],
\end{align}
where $a_{i,j}$ and $a^{\dagger}_{i,j}$ denote the annihilation and creation operator on site $(i,j)$, respectively. The parameters are defined as follows,
\begin{align}
\begin{split}
t^{x}_{i,j}=&-\frac{\hbar^{2}}{2ma^{2}}\frac{\sqrt{g_{i,j}}g_{i,j}^{xx}}{(g_{i+1,j}g_{i,j})^{1/4}},\\
t^{y}_{i,j}=&-\frac{\hbar^{2}}{2mb^{2}}\frac{\sqrt{g_{i,j}}g_{i,j}^{yy}}{(g_{i,j+1}g_{i,j})^{1/4}},\\
u_{i,j}=&\frac{\hbar^{2}}{2ma^{2}}\left(\frac{\sqrt{g_{i-1,j}}}{\sqrt{g_{i,j}}}g_{i-1,j}^{xx}+g_{i,j}^{xx}+\frac{a^{2}}{b^{2}}\frac{\sqrt{g_{i,j-1}}}{\sqrt{g_{i,j}}}g_{i,j-1}^{yy}+\frac{a^{2}}{b^{2}}g_{i,j}^{yy}\right)-\frac{\hbar^{2}\kappa}{8m}.
\end{split}
\end{align}
Substituting the metric tensor of the \Poincare half-plane and the \Poincare disk into Eq.(\ref{2ndH}), one could obtain Eq.(5) and Eq.(7) in the main text.

\subsection{Supplementary Note 2: The solution to the Schr\"odinger equation on a \Poincare disk}
For the \Poincare disk,  the metic tensor is written as
\begin{align}
g^{\rho\rho}=g_{\rho\rho}^{-1}=\frac{\left(1-\kappa\rho^{2}\right)^{2}}{4},\quad g^{\theta\theta}=g_{\theta\theta}^{-1}=\frac{\left(1-\kappa\rho^{2}\right)^{2}}{4\rho^{2}},
\end{align}
and,
\begin{align}
\sqrt{g}=\sqrt{g_{\rho\rho}g_{\theta\theta}}=\frac{4\rho}{\left(1-\kappa\rho^{2}\right)^{2}}.
\end{align}
Thus, the Schr\"odinger equation is written as
\begin{align}
-\frac{\hbar^{2}}{2m}\left[\frac{(1-\kappa\rho^{2})^{2}}{4\rho}\frac{\partial}{\partial \rho}\rho\frac{\partial}{\partial \rho}+\frac{(1-\kappa\rho^{2})^{2}}{4\rho^{2}}\frac{\partial^{2}}{\partial \varphi^{2}}+\frac{\kappa}{4}\right]\Psi(\rho,\theta)=E\Psi(\rho,\theta),
\end{align}
and the normalization is
\begin{align}
\int_{0}^{\infty}\frac{4\rho d\rho}{(1-\kappa\rho^{2})^{2}}\int_{0}^{2\pi}d\theta\Psi^{*}(\rho,\theta)\Psi(\rho,\theta)=1.
\end{align}
The main text has discussed the solution to the Schr\"odinger equation when the boundaries are horocycles. Here, we consider boundaries that are concentric circles about the origin. Under this condition, the wave function can be factorized as $\Psi(\rho,\theta)=\frac{1}{\sqrt{2\pi}}\varphi(\rho)e^{in\theta}$ in the polar coordinates. To simplify notations, we adopt the unit $\hbar=2m=1$, and the Schr\"odinger equation for the radical part is then written as
\begin{align}
\label{radicalSch}
\left[\rho\frac{\partial}{\partial \rho}\rho\frac{\partial}{\partial \rho}+\frac{\rho^{2}}{(1-\kappa\rho^{2})^{2}}(1+4E/\kappa)\right]\varphi(\rho)=n^{2}\varphi(\rho).
\end{align}
\begin{figure}
\centering
\includegraphics[width=0.8\textwidth]{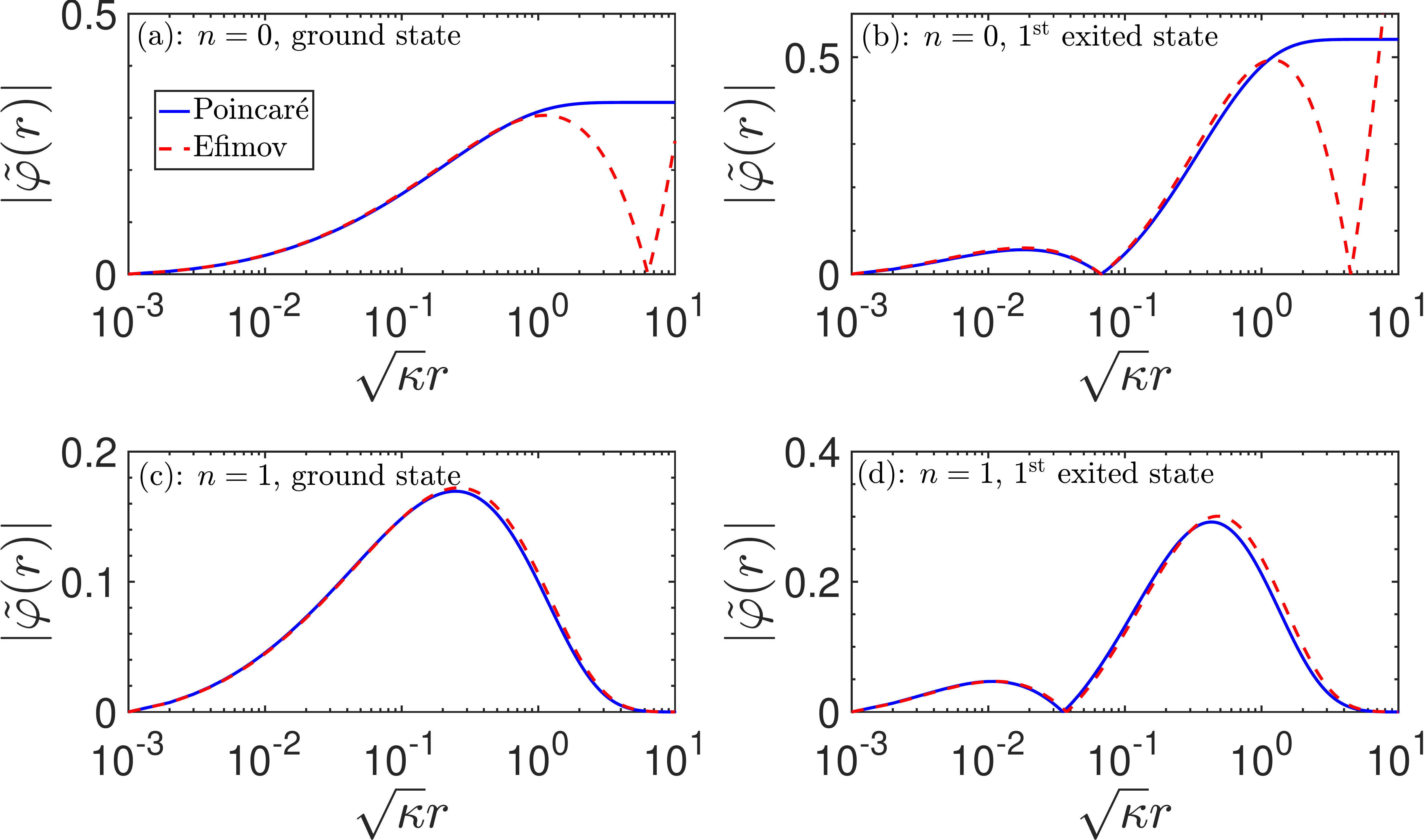} 
\caption{ Comparisons of the solutions to Eq.(\ref{radicalSch1}) (solid curves) and that  to Eq.(\ref{radicalSch2}) (dashed curves). In our calculations, we use the concentric boundary condition $\varphi(r_{0})=0$ with $r_{0}=0.001/\sqrt{\kappa}$ and $\varepsilon=1$. $n$ is the quantum number of the angular momentum. Both the ground (a,c) and excited states (b,d) in the radial direction have been shown. 
\label{diskwf}}
\end{figure}

If we define a new variable, $\varepsilon \sqrt{\kappa}r=-\ln(\sqrt{\kappa}\rho)$ ($\varepsilon$ is arbitrary and dimensionless number), 
the connection between Eq.(\ref{radicalSch}) and Efimov physics becomes clear. In this new coordinate, Eq.(\ref{radicalSch}) becomes 
\begin{align}
\label{radicalSch1}
\left[\frac{\partial^{2}}{\kappa\varepsilon^{2}\partial r^{2}}+\frac{1/4+E/\kappa}{\sinh^{2}(\varepsilon\sqrt{\kappa} r)}\right]\tilde\varphi(r)=n^{2}\tilde\varphi(r)
\end{align}
with the normalization 
\begin{align}
\int_{0}^{\infty}\frac{dr}{\sqrt{\kappa}\sinh^{2}(\varepsilon\sqrt{\kappa}r)}\tilde\varphi^{*}(r)\tilde\varphi(r)=1.
\end{align}
The solutions to Eq.(\ref{radicalSch1}) are written as
\begin{align}
\tilde\varphi(r)\propto P_{\ell_{+}}^{m}\left(\coth(\varepsilon\sqrt{\kappa}r)\right)+{\cal C} P_{\ell_{-}}^{m}\left(\coth(\varepsilon\sqrt{\kappa}r)\right)
\end{align}
where $\ell_{\pm}=-1/2\pm i\sqrt{E/\kappa}$ and  $P_{\ell}^{m}(z)$ is the associated Legendre function. Since $P_{\ell}^{m}=P_{-\ell-1}^{m}$, the choice of $\ell_{\pm}$ here is irrelevant \cite{chaosonpseudosphere}. 
We thus take ${\cal C}=0$ in our calculations.
In the limit that $\varepsilon r\to0$, i.e., $\rho\to1$ (the boundary of the \Poincare disk), Eq.(\ref{radicalSch1}) can be well approximated by 
\begin{align}
\label{radicalSch2}
\kappa\left(\frac{\partial^{2}}{\partial r^{2}}+\frac{1/4+E/\kappa}{r^{2}}\right)\tilde{\varphi}(r)=\varepsilon n^{2}\tilde{\varphi}(r),
\end{align}
and the normalization becomes
\begin{align}
\int_{0}^{\infty}\frac{dr}{\kappa^{3/2}r^{2}}\tilde\varphi^{*}(r)\tilde\varphi(r)=1.
\end{align}
The solution to Eq.(\ref{radicalSch2}) is written as
\begin{align}
\tilde\varphi(r)\propto \sqrt{r}K_{i\sqrt{E/\kappa}}\left(\varepsilon\sqrt{\kappa}r\right),
\end{align}
where $K_{a}(y)$ denotes the modified Bessel function of the second kind.

In other words, if the concentric boundary is close to the edge of the \Poincare disk ($\rho\rightarrow 1$ or $r\to0$), the wave functions approach Efimov states near the boundary. This can be understood from the fact that horocycles near the edge of the \Poincare disk approach concentric circles about the origin. However, moving away from the boundary circle, the solution deviates more and more significantly from Efimov states, as shown in Fig. \ref{diskwf}. For a small $\rho$ or large $r$, the eigenstates are no longer Efimov states. 

\subsection{Supplementary Note 3: Shifting funneling mouths}
\begin{figure}
\centering
\includegraphics[width=0.9\textwidth]{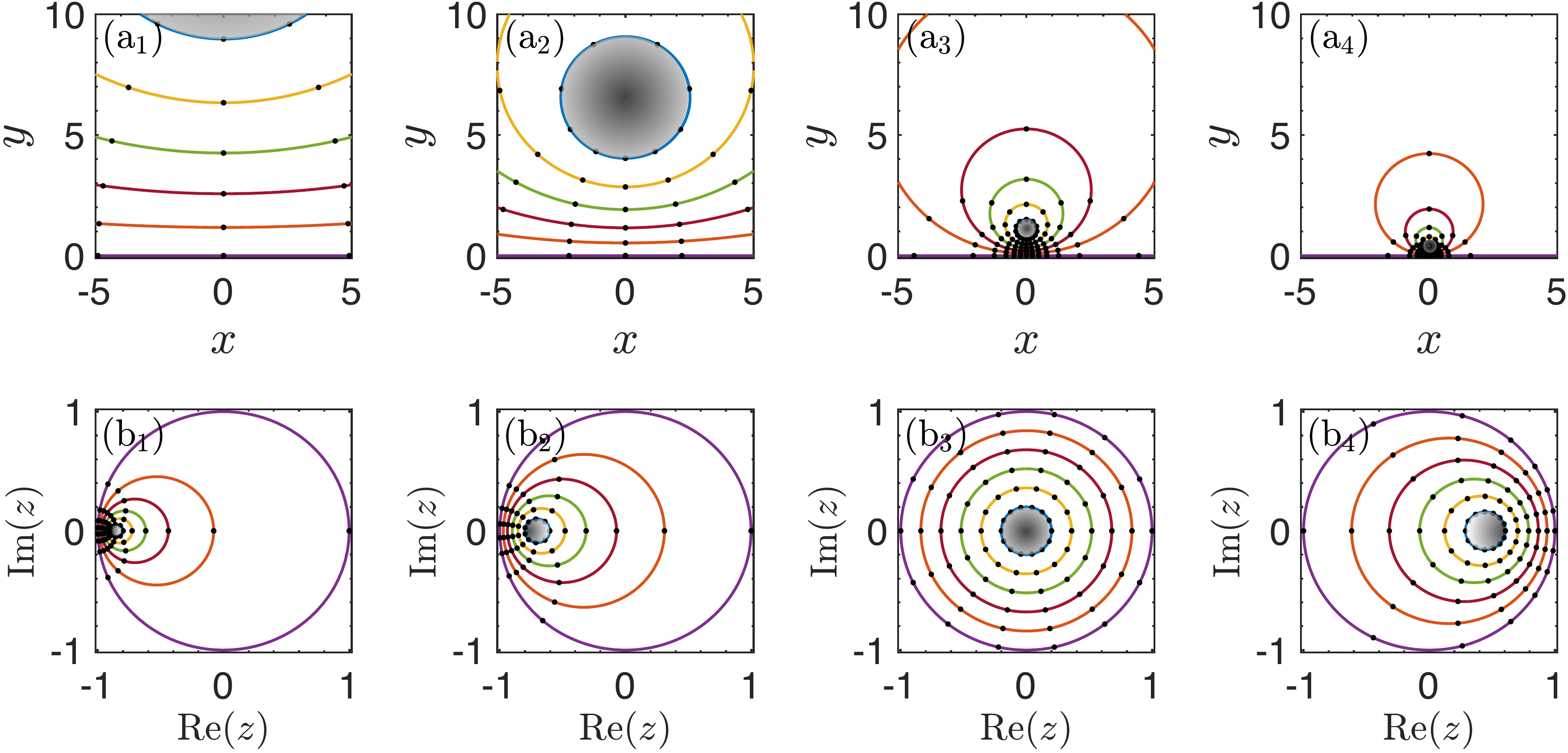} 
\caption{The shifting of the funneling mouths on \Poincare half-plane and \Poincare disk by changing the boundary condition. The funneling mouths are denoted by the shaded regions. (${\rm a_{1}}$-${\rm a_{4}}$): funneling mouths on the \Poincare half plane. (${\rm b_{1}}$-${\rm b_{4}}$): funneling mouths on the \Poincare disk.  The curves indicate how to deform the lattice discussed in the main text so as to shift the funneling mouth to a desired location on these two hyperbolic spaces. The lattice sites are symbolized by the black dots. The parameter are taken as follows : $\beta=-1.3$ (${\rm a_{1},b_{1}}$); $\beta=-0.9$ (${\rm a_{2},b_{2}}$); $\beta=0$ (${\rm a_{3},b_{3}}$); $\beta=0.5$ (${\rm a_{4},b_{4}}$). 
\label{funnelmouth}}
\end{figure}

The funneling mouths on the \Poincare half-plane and the \Poincare disk can be shifted by changing the boundary in the continuous space and rearranging locations of lattice sites correspondingly. To this end, we resort to the \Mobius transformation on the \Poincare disk,
\begin{align}
\label{diskmobius}
z'=\frac{z\cosh(\beta)+\sinh(\beta)}{z\sinh(\beta)+\cosh(\beta)},
\end{align}
where $z$ and $z'$ are complex numbers and symbolize the points on \Poincare disk, and $\beta\in(-\infty,\infty)$.  The mapping defined in Eq. (\ref{diskmobius}) preserves the interior of the \Poincare disk. Invoking Eq.(\ref{diskmobius}), we could map the original location of the funneling mouth to any other point on the \Poincare disk. Accordingly, all lattice sites are shifted by the \Mobius transformation specified by the same $\beta$. In other words, the position of the funneling mouth can be continuously tuned by changing $\beta$, as illustrated in Fig.\ref{funnelmouth}(${\rm b_{1}}$-${\rm b_{4}}$). Similarly, the funneling mouth on the \Poincare half-plane is also continuously tuned via the \Mobius transformation Eq.(\ref{diskmobius}) followed by the Cayley transformation defined in Eq.(10) of the main text, as shown in Fig.\ref{funnelmouth}(${\rm a_{1}}$-${\rm a_{4}}$). 

The funneling mouth shown in Fig. 3  of the main text corresponds to $\beta\to-\infty$. Thus the funneling mouth on the \Poincare half-plane locates at $y=\infty$ and the counterpart on the \Poincare disk is a horocycle. In the main text, we choose the boundary as a short range cut-off $y=y_0$ with $y_0>0$ being a constant. By tuning the parameter $\beta$ from $-\infty$ to $0$ the horizontal boundary $y=y_0$ on the \Poincare half plane is bent and the funneling mouth gradually moves from $y=\infty$ to finite $y$. Their counterparts on the \Poincare disk move towards the center of the disk. This process is depicted in  Fig.\ref{funnelmouth} (${\rm a_{1}}\to{\rm a_{3}}$) and (${\rm b_{1}}\to{\rm b_{3}}$). Here, the funneling mouths are encircled by the blue curves. When the funneling mouth locates at $(0,1)$ on \Poincare half-plane, the counterpart on \Poincare disk is at the center, as shown in Fig.\ref{funnelmouth} (${\rm a_{3},b_{3}}$). If $\beta$ further increases,  the funneling mouth on \Poincare disk will leave the center and move to the other side of the center, as shown in Fig.\ref{funnelmouth} (${\rm a_{4},b_{4}}$).  In this process, all lattice sites are relocated by the corresponding \Mobius transformations and the tunnelings between the nearest neighbor sites remain unchanged, as illustrated by the black dots in Fig. \ref{funnelmouth}.

\end{document}